# Jet électromagnétique vers des applications de caractérisation

A. Ghaddar[1], S. Touil[1], B. Bayard[1] and B. Sauviac[1]

[1] *Université de Lyon, Université Jean Monnet-Saint-Etienne, CNRS, Institut d'Optique Graduate School, Laboratoire Hubert Curien*
*UMR 5516, F-42023, SAINT-ETIENNE, France*
ali.ghaddar@univ-st-etienne.fr

*Résumé* — **Cette étude comparative permet d'évaluer la performance d'un jet électromagnétique à déterminer la réponse électromagnétique des matériaux, sans être dans des conditions habituelles de champ lointain. Dans ce travail, le coefficient de réflexion d'un substrat avec plan de masse, calculé analytiquement, est comparé avec celui déterminé en simulation par le jet électromagnétique. Les résultats présentent un accord satisfaisant après avoir calibré le coefficient de réflexion simulé au point focal du jet. Ainsi, nous avons pu retrouver, par une mesure locale, des données cohérentes avec une mesure classique de type espace libre.**

## I. INTRODUCTION

Le concept de jet électromagnétique a connu récemment des progrès croissants en raison de ses applications en détection et en imagerie allant du domaine optique aux fréquences microondes. Ce concept a été proposé pour la première fois dans le domaine optique (jet photonique) en 2004 par Z. Chen et *al*. [1]. Dans ce travail, ils ont observé que l'interaction de l'onde optique plane avec une nanoparticule donne une onde focalisée en champ proche (dans l'ombre de la nanoparticule) avec une largeur à mi-hauteur d'une demi-longueur d'onde. Cette onde focalisée présente une forte intensité du champ avec une propagation sur quelques longueurs d'ondes. Ils ont montré également que ce nano jet permet d'améliorer la rétrodiffusion des objets ayant des tailles faibles devant la longueur d'onde, cela en fait par conséquent un bon moyen pour la détection et l'imagerie. Ce type de jet électromagnétique a également été étendu aux basses fréquences en considérant des embouts de différentes formes géométriques (elliptique, cylindrique, ...) à la sortie d'un guide d'onde ou d'une antenne [2], [3]. Quel que soit le domaine fréquentiel, le jet électromagnétique reste encore peu utilisé dans des applications de type détection ou imagerie. Ces deux applications peuvent être réalisées en analysant directement l'évolution du coefficient de réflexion. Ainsi, nous visons dans ce travail d'étendre le domaine d'application du jet électromagnétique en l'utilisant pour extraire des informations sur le matériau à travers le post-traitement du coefficient de réflexion à l'aide des équations de Maxwell. A titre d'exemple, nous essayerons de déterminer la permittivité du substrat (placée au point focal du jet) à partir de sa réponse électromagnétique en réflexion détectée par le jet.

## II. RÉSULTATS ET DISCUSSION

### A. *Jet électromagnétique*

Le jet électromagnétique présenté dans cette étude est une antenne cornet chargée de Téflon prolongée par un cylindre (voir Fig. 1.). La charge en Téflon est constituée d'une partie transitoire à l'entrée de l'antenne en permettant de minimiser les réflexions dues au changement de milieu (Air/Téflon), et d'un embout cylindrique de la hauteur h=2,4 cm et du rayon R=0,9 cm à la sortie de l'antenne. Comme illustré par la carte du champ électrique E, cet embout cylindrique permet de focaliser l'onde de fréquence 30 GHz à une distance de 5 mm après la terminaison de l'embout (distance entre le point central de la tâche la plus intense du champ et la terminaison de l'embout).

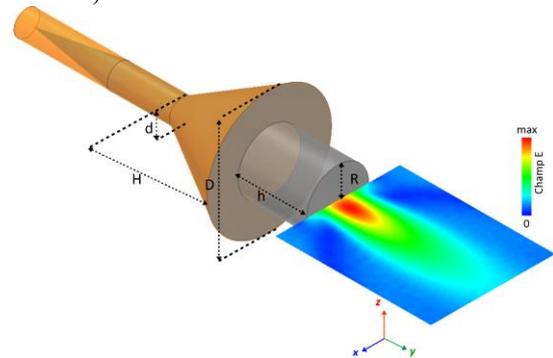

Fig. 1. Représentation de l'antenne cornet de dimensions d=0,82 cm, H=2,8 cm et D=4 cm chargée de Téflon prolongée en espace libre par un embout cylindrique de dimensions h=2,4 cm et R=0,9 cm. La carte du champ électrique a été tracée à la fréquence 30 GHz dans le plan xy.

Pour comprendre profondément le principe de fonctionnement du jet électromagnétique, nous avons tracé la carte du champ électrique à l'intérieur et à l'extérieur de l'embout en présence d'un matériau au point focal. Comme montre la figure 2, l'onde électromagnétique rencontre deux milieux différents à la sortie de l'antenne cornet, l'embout cylindrique du Téflon au centre entouré de l'air aux extrémités. La partie de l'onde submergée dans l'air présente une vitesse de propagation plus rapide que la partie de l'onde se propageant dans le Téflon. Le front d'onde à la sortie de l'antenne cornet a une forme concave et l'onde redevient sous forme convexe à l'extrémité de l'embout après l'avoir traversé. Cette déformation est expliquée par la différence de vitesse de propagation entre l'air et le Téflon. Ensuite, la réfraction de l'onde vers l'air permet d'amplifier la convexité de l'onde en nous donnant à la fin une tache plus intense à 5 mm de la terminaison de l'embout. Ainsi, la différence de vitesse de propagation entre l'air et le Téflon est à l'origine de phénomène de focalisation de notre jet électromagnétique. Cependant, l'onde qui se propage aux extrémités présente une intensité négligeable devant la tache centrale en raison de perte de l'éloignement. L'onde focalisée semble être moins déformée avec une concentration du champ centrale qui peut être considérée approximativement plane. Pour observer la forme de plusieurs longueur d'onde au lieu qu'une seule, nous avons placé au point focal un matériau sans perte et plus réfringent de l'indice n= 3,15. A l'intérieur du matériau l'onde électromagnétique a une longueur d'onde trop petite avec des fronts remplisses parfaitement la condition d'onde plane. Ce qui nous permet d'avoir des résultats comparables à ceux issus des équations de Maxwell dans le cadre du formalisme de l'onde plane.

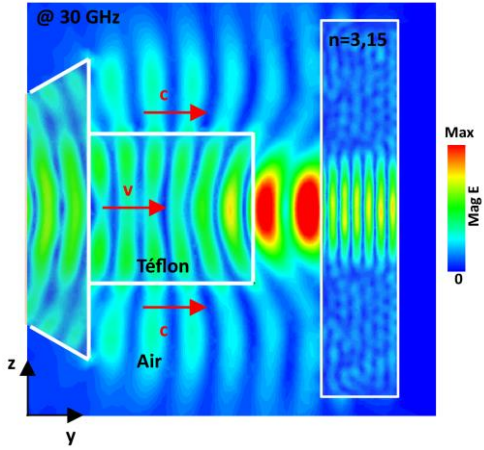

Fig. 2. Carte du champ E en vue de coupe représentant l'évolution de l'onde à travers l'embout et à l'intérieur du matériau placé à 5 mm de la terminaison de l'embout.

A l'aide du jet électromagnétique introduit ci-dessus, nous allons étudier la réponse électromagnétique d'un matériau simple composé d'un substrat avec plan de masse en face arrière. Les résultats de simulation que l'on va présenter par la suite ont été obtenus en plaçant la surface supérieure du substrat au point focal du jet perpendiculairement à l'axe y. Par ailleurs, le coefficient de réflexion obtenu en simulation nécessitera un calibrage post-simulation afin de minimiser les pertes et les erreurs résultant de l'enchaînement entre l'entrée et la sortie de l'antenne cornet. La calibrage du coefficient de réflexion est basée sur le calibrage de type Reflect/Match [2]. En simulation, la condition Reflect correspond au coefficient de réflexion $S_{11(reflect)}$ en présence d'une surface réfléchissante électrique parfaite PEC au point focal du jet. Quant à la condition Match, le coefficient de réflexion $S_{11(match)}$ a été simulé en mettant au point focal les conditions d'absorption PML (Perfect Matched Layer). Le coefficient de réflexion corrigé $S_{11(cor)}$ peut donc être obtenu par la relation suivante :

$$S_{11\,(cor)} = \frac{S_{11} - S_{11\,(match)}}{S_{11\,(match)} - S_{11\,(reflect)}} \quad (1)$$

La performance du jet à la détermination de la réponse en réflexion d'un substrat avec plan de masse est évaluée en la comparant avec le résultat analytique détaillé dans le paragraphe suivant.

*B. Modélisation théorique*

La réponse électromagnétique des systèmes multicouches de différents indices peut être traitée analytiquement en utilisant les équations de Maxwell. Pour déterminer l'expression analytique du coefficient de réflexion associée à un substrat avec plan de masse, il nous faudra considérer un système constitué de trois milieux : l'espace libre au-dessus du substrat, le substrat d'indice n et le plan métallique sous le substrat. Dans ce cas, le coefficient de réflexion $S_{11}$ est défini selon la relation (2). En plus, l'épaisseur du substrat peut être déterminée à partir du coefficient de réflexion en utilisant la relation (3).

$$S_{11} = \frac{1 - n - (n+1)e^{-2ikd}}{1 + n + (n-1)e^{-2ikd}} \quad (2)$$

Dans cette expressions n, k et d correspondent respectivement à l'indice de réfraction du substrat, au vecteur de propagation dans le substrat et à l'épaisseur du substrat.

*C. Comparaison analytique/simulation*

Le module et la phase du coefficient de réflexion en fonction de l'épaisseur obtenus à la fois analytiquement et en simulation via le jet électromagnétique sont présentés sur la figure 2.

Les résultats présentés ont été obtenus en considérant un substrat d'indice n = 2,37-j0,2. Comme montré sur la figure 3, le module du coefficient de réflexion $S_{11}$ obtenu directement en simulation sans calibrage présente une évolution similaire à celui calculé analytiquement à l'exception d'un décalage de 0,2 entre eux. Ce décalage est quasiment supprimé après avoir calibré le paramètre $S_{11}$ en se basant sur la relation (1). En revanche, la phase de $S_{11}$ reste très proche du calcul analytique, que ce soit avec ou sans calibrage. Cette cohérence entre la simulation et l'approche analytique confirme que le jet électromagnétique permet de détecter correctement la réponse du matériau. Cela peut ouvrir la voie au jet électromagnétique vers des nouvelles applications en caractérisation des matériaux.

*D. Expérimentation*

La validation expérimentale de la fiabilité du jet à détecter correctement la réponse du matériau a été menée sur un échantillon de FR4 Epoxy de permittivité ε = 4,4(1- j 0,04) et d'épaisseur 1,6 mm. L'échantillon de FR4 Epoxy sous test présenté sur la figure 4 est constitué de trois parties : un plan métallique, un substrat avec plan de masse et un substrat sans plan de masse. Le plan métallique ajouté sur l'échantillon est utilisé comme condition Reflect pour calibrer les mesures effectuées sur le substrat de FR4 Epoxy. L'ajout de ce plan métallique sur le même échantillon sous test permettra d'obtenir une condition de calibrage plus rigoureuse, tout cela en évitant les erreurs de positionnement introduites lorsque la condition Reflect et les mesures sur l'échantillon sont faites séparément. Les mesures ont été prises en plaçant l'échantillon au point focal du jet électromagnétique. Afin de retrouver expérimentalement le point focal, on déplace le plan métallique devant le jet jusqu'à l'obtention d'une position à laquelle le coefficient de réflexion présente une valeur maximale. A l'aide d'un bras robotisé, la caractérisation a été réalisée continument le long de l'échantillon en le déplaçant pas-à-pas perpendiculairement au jet électromagnétique (suivant l'axe z).

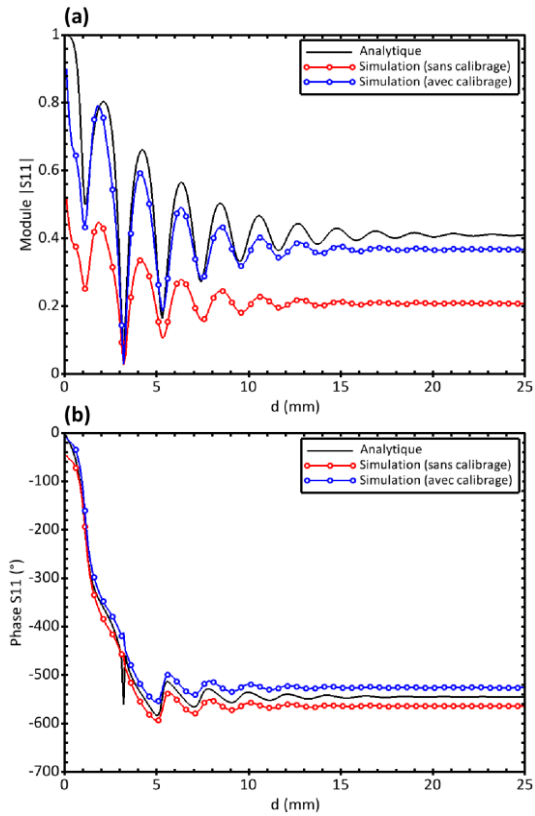

Fig. 3. (a) et (b) sont respectivement le module et la phase du coefficient de réflexion $S_{11}$ en fonction de l'épaisseur du substrat d. Le trait noir représente le résultat obtenu analytiquement, les résultats représentés par des cercles rouges et bleus correspondent respectivement à la simulation sans et avec calibrage.

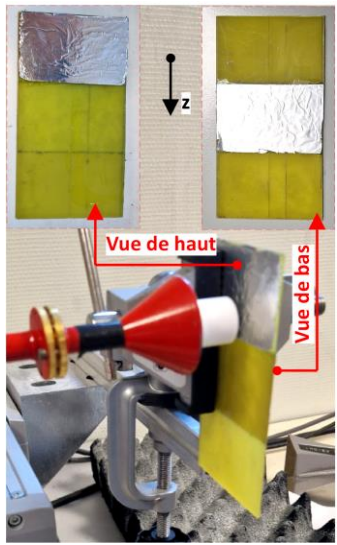

Fig. 4. Photo du banc de caractérisation d'un échantillon de substrat FR4 Epoxy constitué d'une partie de plan métallique, une partie de substrat avec plan de masse et sans plan de masse. En insert, les photos de la face antérieure (vue de haut) et postérieure (vue de bas) de l'échantillon.

L'évolution du module et de la phase du coefficient de réflexion expérimental le long de l'échantillon à la fréquence 30 GHz est présentée sur la figure 5. Le jet électromagnétique a commencé de prendre les mesures sur la partie du plan métallique (à z=0) et il a terminé sur le substrat sans plan de masse (à z=100 mm) après être passé au milieu sur la partie du substrat avec plan de masse. Entre 0 et 19.5 mm, le module et la phase du coefficient de réflexion du plan métallique prennent respectivement des valeurs autour de 0 dB et -180 °. Entre z=19,5 et 30 mm, le module du coefficient de réflexion présente une chute et puis une ascension en reprenant une stabilité autour de -0,98 dB. La phase a également montré dans cette zone une transition de -180° à -77°. Cette évolution est associée à la transition du jet du plan métallique au substrat avec plan de masse. Physiquement, la diminution du module $S_{11}$ dans cette zone peut être expliquée par le fait que le champ électromagnétique généré par le jet recouvert une partie sur le plan métallique et d'autre partie sur le substrat avec plan de masse, ce qui engendre un retour de l'onde orientée en dehors du jet en raison de l'avancement de phase d'une partie par rapport à l'autre. Ainsi, le gap entre ces deux états stables peut être lié au diamètre du champ électromagnétique généré par le jet au point focal (voir la figure 1). Le même phénomène est observé lorsque le jet traverse le substrat avec plan de masse et rentre dans la partie du substrat sans plan de masse. Finalement, le module et la phase mesurés sur cette dernière partie présente des valeurs respectivement autour de -5.2 dB et -200°. Comme illustré sur la figure 5, le coefficient de réflexion expérimental de chaque type de matériau est comparé avec la théorie représentée par des courbes droites en traits. Cette comparaison montre un accord satisfaisant entre la mesure et les valeurs attendues à 30 GHz. Ce qui confirme la capacité du jet à détecter correctement la réponse du matériau.

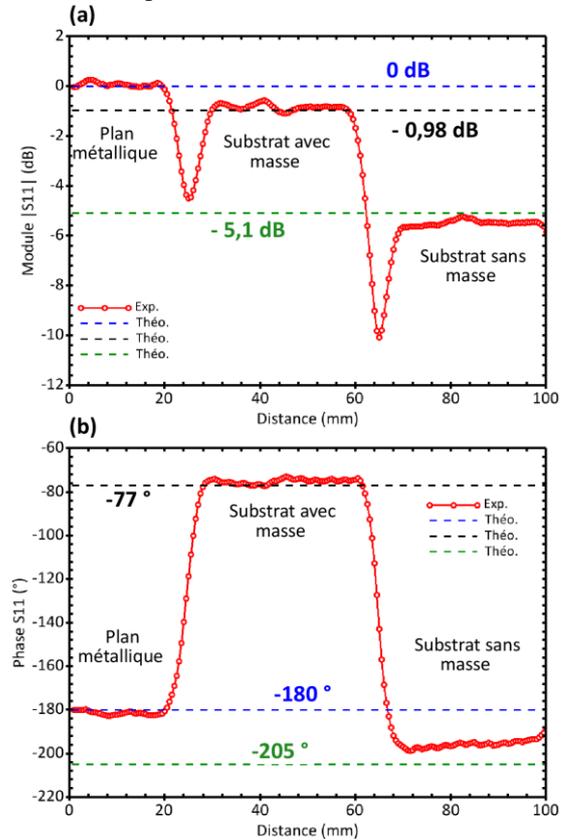

Fig. 5. Module (a) et la phase (b) du coefficient de réflexion $S_{11}$. La courbe en cercle rouge représente le résultat expérimental (Exp.). Les courbes droites en traits représentent les valeurs théoriques à 30 GHz (Théo.).

*E. Exploitation de la méthode pour la détermination de grandeurs caractéristiques des échantillons*

Nous essayons dans ce paragraphe d'utiliser les mesures de paramètre $S_{11}$ effectuées sur l'échantillon de FR4 Epoxy pour but de retrouver expérimentalement la valeur de sa permittivité. De ce fait, une méthode de l'extraction de la permittivité basée sur le paramètre $S_{11}$ a été appliquée [4], [5]. Cette méthode se repose sur deux mesures réalisées à la fois sur le substrat avec plan de masse et sur le substrat sans plan de masse. La permittivité est calculée à partir de l'impédance du matériau $Z_{sans}$ seul et du matériau avec plan de masse $Z_{avec}$ selon l'expression suivante :

$$\varepsilon = \frac{Z_0^2}{Z^2} + \sin^2\theta \qquad (3)$$

$$Z^2 = \frac{Z_0\, Z_{avec}\, Z_{sans}}{Z_0 + Z_{avec} - Z_{sans}} \qquad (4)$$

$$Z_{avec/sans} = Z_0 \frac{1 + S_{11\,avec/sans}}{1 - S_{11\,avec/sans}} \qquad (5)$$

Avec $\theta$, l'angle incidente de l'onde électromagnétique. Dans notre cas, nous considérons le jet génère les ondes électromagnétiques en incidence normale ($\theta=0$). Et $Z_0$, l'impédance dans l'air.

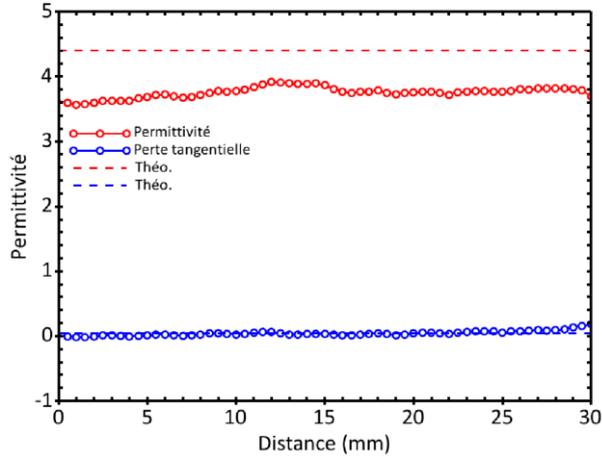

Fig. 6. Permittivité et perte tangentielle expérimentales représentées respectivement par des cercles rouges et bleus en comparaison avec les valeurs attendues représentées par les courbes droites en trait.

La permittivité expérimentale a été obtenue en utilisant la mesure réalisée sur le substrat avec plan de masse et celle sur le substrat sans plan de masse, plus précisément, les mesures situées dans la gamme 30-60 mm et 70-100mm, respectivement. La permittivité et la perte tangentielle étalonnées par la méthode sont présentées sur la figure 6. La valeur expérimentale moyenne de la permittivité complexe présente un écartement de $0,65(1 + j\, 4.10^{-3})$ par rapport à la valeur correcte qui vaut $\varepsilon = 4,4(1 - j\, 0,04)$. Cet écartement provient peut-être de la mesure du paramètre $S_{11}$ effectuée sur le substrat sans plan de masse qui présente une petite différence comparativement à la valeur théorique (voir figure 5). Cette différence par rapport à la valeur exacte peut donc être réduit en améliorant la mesure correspondante. On note que l'échantillon doit être parfaitement positionné devant le jet électromagnétique. Sinon, une faible inclinaison peut entrainer des erreurs sur les mesures car le jet est très sensible à la détection.

## III. CONCLUSION

Un bon accord a été obtenu entre la réponse d'un matériau détectée par le jet et la réponse théorique, calculée analytiquement en utilisant le formalisme du champ lointain. Une expérimentation a été réalisée ensuite afin de valider pratiquement cette approche théorique. Cette technique peut donc être utilisée pour extraire des informations complémentaires sur l'échantillon testé.

Ainsi, l'utilisation du jet électromagnétique ne serait plus limitée aux applications de la détection et l'imagerie, mais pourrait être exploitée également pour des activités de caractérisation.

## REFERENCES

[1] Z. Chen, A. Taflove, et V. Backman, « Photonic nanojet enhancement of backscattering of light by nanoparticles: a potential novel visible-light ultramicroscopy technique », *Opt. Express*, vol. 12, nº 7, p. 1214, 2004, doi: 10.1364/OPEX.12.001214.
[2] H. Hyani, « Jet électromagnétique 3D : détection, imagerie et contrôle non destructif dans des structures opaques. », p. 141.
[3] B. Ounnas, B. Sauviac, Y. Takakura, S. Lecler, B. Bayard, et S. Robert, « Single and Dual Photonic Jets and Corresponding Backscattering Enhancement With Tipped Waveguides: Direct Observation at Microwave Frequencies », *IEEE Trans. Antennas Propagat.*, vol. 63, nº 12, p. 5612-5618, déc. 2015, doi: 10.1109/TAP.2015.2491328.
[4] R. A. Fenner, E. J. Rothwell, et L. L. Frasch, « A comprehensive analysis of free-space and guided-wave techniques for extracting the permeability and permittivity of materials using reflection-only measurements », *Radio Science*, vol. 47, nº 01, p. 1-13, 2012, doi: 10.1029/2011RS004755.
[5] L. Pometcu, « Matériaux et forme innovants pour l'atténuation en hyper fréquences », Université de Rennes 1, Rennes, 2016.